\documentstyle[12pt,aps]{revtex}

\widetext
\draft
\tighten
\begin{document}

\title{\normalsize{\rm{\bf GENERALIZATIONS OF THE DIRAC EQUATION
AND THE MODIFIED BARGMANN-WIGNER FORMALISM}}\thanks{Some parts of this work have
been presented at {\it the Second International Workshop
"Graphs-Operads-Logic, Parallel Computation, Mathematical Physics", May
6-16, 2002, FESC-UNAM, Cuautitl\'an, M\'exico} and  the {\it 6th
International  Conference on Clifford Algebras and Their Applications in
Mathematical Physics, May 20-25, 2002,  Tennessee Technological
University, Cookeville, TN, USA}.}}

\author{{\bf Valeri V. Dvoeglazov}}

\address{{\rm Universidad de Zacatecas, Apartado Postal 636,
Suc. UAZ\\Zacatecas 98062, Zac., M\'exico\\
E-mail: valeri@ahobon.reduaz.mx\\
URL: http://ahobon.reduaz.mx/\~\,valeri/valeri.htm}}

\date{June 16, 2002}

\maketitle

\bigskip

\begin{abstract}
{\bf Abstract.}
We present various generalizations of the Dirac formalism.
The different-parity solutions
of the Weinberg's $2(2J+1)$-component equations
are found. On this basis, generalizations of the Bargmann-Wigner
(BW) formalism are proposed. Relations with modern physics constructs are
discussed.  \end{abstract}

\bigskip

\section{Introduction}

In  this work I am going to discuss the following matters:

\begin{itemize}

\item
Generalizations of the Dirac equation. Why? Who? How far?

\item
The theory in the $(J,0)\oplus (0,J)$ representation of the Lorentz group
(the Weinberg $2(2J+1)$ theory).

\item
A lot of antisymmetric tensor (AST) fields. The Proca theory is
generalized.

\item
(The standard Bargmann-Wigner $(2J+1)$-formalism). $\Leftrightarrow$ (The
Proca-Duffin-Kemmer formalism.)

\item
(Modifications of the Bargmann-Wigner formalism.) $\Leftrightarrow$
(Modifications of the Proca theory.)

\item
Conclusions. What further?

\end{itemize}

What is the purpose of my research?  I am sure that
the modern physics constructs have much deeper relations with
the space-time and discrete symmetries than it was believed before.

\section{Generalizations of the Dirac formalism}

First of all, I am going to present some preliminary material
(perhaps, not so known as it deserves). The equations below can be derived
from the first principles and they are used for
modifications of the Bargmann-Wigner formalism.

\bigskip

\subsection{The Tokuoka-Sen Gupta-Fushchich formalism}

It is based on the equation~\cite{g1,g2,g3,g3a}:
\begin{equation}
[i\gamma_\mu \partial_\mu +m_1 + m_2 \gamma^5] \Psi =0
\end{equation}
\begin{itemize}

\item
If $m_1^2\neq m_2^2$ it was claimed~\cite{g1} that this is simply the change of the representation
of $\gamma$'s.

\item
Physical consequences are:

\begin{enumerate}

\item
The equation can describe bradyonic, massless and tachyonic particles depending on the choice of
the parameters $m_1$ and $m_2$, ref.~\cite{g3}.

\item
The equation we get under the charge conjugation
operation is {\it not} the same equation; the sign before $\gamma^5$ term
is changed to the opposite one~\cite{g1}, provided that $m_1, m_2 \in\Re
e$.

\item
It is impossible to construct a Lagrangian unless
we introduce the second field, e.~g., $\Psi^c$.

\item
In the massless case the solutions are {\it no} longer eigenstates
of the $\gamma_5$ operator: neither they are eigenstates of the helicity
operator of the $(1/2,0)\oplus (0,1/2)$ representation~\cite{g1}.

\item
The commutation relations are quite unusual~\cite{g2},
e.g., for massless particles $m_1= \pm m_2$ one has
$\{\Psi_\sigma,\overline\Psi_{\sigma^\prime}^c\}\neq 0$,
$\{\Psi_\sigma,\Psi_{\sigma^\prime}\}\neq 0$, but
$\{\Psi_\sigma,\overline\Psi_{\sigma^\prime}\}=0$.
However, even in the massive case we  can have nonlocality, i.e. the
presence of the even Pauli-Jordan function~\cite{dvo-hj}.

\end{enumerate}

\item
Fushchich~\cite{Fush} generalized the formalism
even further in 1970-72, and, in fact, he connected it with the
Gelfand-Tsetlin-Sokolik idea~\cite{gts} of the 2-dimensional
representation of the inversion group.\footnote{This type of theories is
frequently called  the Wigner-type. However, the Wigner lectures turn
out to have been presented later (1962-64).}

\item
I derived the above parity-violating
equation~\cite{g3a} (and its charge-conjugate) by
the Sakurai-Gersten method from
the first principles, see the Appendix A.

\end{itemize}

\subsection{The Barut Formalism}

It is based on the equation~\cite{barut,Wilson}:
\begin{equation}
[i\gamma_\mu \partial_\mu +\alpha_2 {\partial_\mu \partial_\mu \over  m} +
\ae]
\Psi =0\,.
\end{equation}
It was re-derived from the first principles in~\cite{Dvo,afd-dvo}.
\footnote{The Ryder relation
between zero-momentum left- and right- 2-spinors has been
generalized~\cite{Dvo}:
\begin{equation} \phi_L^h (\overcirc{p}^\mu) = a
(-1)^{1/2-h} e^{i(\vartheta_1 +\vartheta_2)} \Theta_{[1/2]} [\phi_L^{-h}
(\overcirc{p}^\mu )]^\ast +b e^{2i\theta_h} \Xi_{[1/2]}^{-1} [\phi_L^h
(\overcirc{p}^\mu )]^\ast\,.
\end{equation}
see for the notation therein.
As a result, in the Majorana representation
we have different equations for real and imaginary parts of
the field function:
\begin{mathletters}
\begin{eqnarray}
\left [ a {i\gamma_\mu \partial_\mu \over m} +b +1 \right ] \Psi_1 (x^\mu)
&=&0\,,\\
\left [ a {i\gamma_\mu \partial_\mu \over m} -b +1 \right ]
\Psi_2 (x^\mu) &=&0\,.  \end{eqnarray} \end{mathletters} }

\begin{itemize}

\item

It represents a theory with  the conserved current that is linear in the generators of
the 4-dimensional representation of the $O(4,2)$ group.

\item
Instead of 4 solutions it has 8 solutions with the correct
relativistic dispersion
$E=\pm \sqrt{{\bf p}^2 +m_i^2}$; and, in fact, describes
{\it two} mass states
$m_\mu = m_e (1+{3\over 2\alpha})$, $\alpha$ is the fine structure
constant, provided that the certain physical condition is imposed on the
$\alpha_2$ parameter~\cite{barut}.

\item
One can also generalize the formalism to include the third state,
$\tau$-lepton, see refs.~[8d,11].

\item
Barut also indicated the possibility of  including $\gamma^5$ term.
For instance, the equation can look something like this:
\begin{equation}
[i\gamma_\mu \partial_\mu +a + b\Box +\gamma^5 (c+d\Box)]\psi =0\,,
\end{equation}
which cannot yet be factorized as a product of
two Dirac equations with different masses.

\end{itemize}

\subsection{The Weinberg-Tucker-Hammer (WTH) Formalism}

The basic principles of constructing the theory in the $(J,0)\oplus (0,J)$
representation~\cite{Wein,dvo-rmf} can be seen in Appendix B.

For spin 1 we start from
\begin{equation}
[\gamma_{\alpha\beta} p_\alpha p_\beta +A p_\alpha p_\alpha +Bm^2] \Psi
=0\,, \end{equation} where $p_\mu=-i\partial_\mu$ and
$\gamma_{\alpha\beta}$ are the Barut-Muzinich-Williams covariantly
defined $6\times 6$ matrices given in Appendix B, $\sum_{\mu}^{}
\gamma_{\mu\mu}=0$.  The determinant of $[\gamma_{\alpha\beta} p_\alpha
p_\beta +A p_\alpha p_\alpha +Bm^2]$ is of the 12th order in $p_\mu$.
Solutions with $E^2 -{\bf p}^2 = m^2$, $c=\hbar =1$ can be obtained if and
only if
\begin{equation} \frac{B}{A+1} =1\,,\quad \frac{B}{A-1}=1\,.
\end{equation}
The particular cases are:
\begin{itemize}

\item
$A=0, B=1$ $\Leftrightarrow$ we have  Weinberg's equation for $J=1$
with 3 solutions $E=+\sqrt{{\bf p}^2 +m^2}$, 3 solutions
$E=-\sqrt{{\bf p}^2 +m^2}$, 3 solutions $E=+\sqrt{{\bf p}^2 -m^2}$
and 3 solutions $E=-\sqrt{{\bf p}^2 -m^2}$.

\item
$A=1, B=2$ $\Leftrightarrow$ we have the Tucker-Hammer equation for $J=1$.
The solutions are {\it only}
with $E=\pm\sqrt{{\bf p}^2 +m^2}$.

\end{itemize}

So, the addition of the Klein-Gordon equation may change the physical content even on the {\it free} level.

What are the corresponding equations for antisymmetric tensor field? Proca?
Maxwell? Recently we have shown~\cite{dvo-hpa,dvo-wig}
that one can obtain {\it four} different equations
for antisymmetric tensor fields from the Weinberg $2(2J+1)$
component formalism.  First of all, we note that $\Psi$ is, in fact,
bivector, ${\bf E}_i = -iF_{4i}$, ${\bf B}_i = {1\over 2} \epsilon_{ijk}
F_{jk}$,, or ${\bf E}_i = -{1\over 2} \epsilon_{ijk} \tilde F_{jk}$, ${\bf
B}_i = -i \tilde F_{4i}$, or their combination.  One can single out the
four cases:
\begin{itemize}

\item
$\Psi^{(I)} = \pmatrix{{\bf E} +i{\bf B}\cr
{\bf E} -i{\bf B}\cr}$, $P=-1$, where ${\bf E}_i$ and ${\bf B}_i$ are the
components of the tensor.

\item
$\Psi^{(II)} = \pmatrix{{\bf B} -i{\bf E}\cr
{\bf B} +i{\bf E}\cr}$, $P=+1$, where ${\bf
E}_i$, ${\bf B}_i$ are the components of the tensor.

\item
$\Psi^{(III)} = \Psi^{(I)}$, but (!) ${\bf E}_i$ and ${\bf B}_i$
are the
corresponding vector and axial-vector  components of the
{\it dual} tensor $\tilde F_{\mu\nu}$.

\item
$\Psi^{(IV)} = \Psi^{(II)}$, where ${\bf E}_i$ and ${\bf B}_i$
are the components of the {\it dual} tensor $\tilde F_{\mu\nu}$.

\end{itemize}
The mappings of the WTH equations are:
\begin{mathletters}\begin{eqnarray}
&&\partial_\alpha\partial_\mu F_{\mu\beta}^{(I)}
-\partial_\beta\partial_\mu F_{\mu\alpha}^{(I)}
+ {A-1\over 2} \partial_\mu \partial_\mu F_{\alpha\beta}^{(I)}
-{B\over 2} m^2 F_{\alpha\beta}^{(I)} = 0\,,\label{wth1}\\
&&\partial_\alpha\partial_\mu F_{\mu\beta}^{(II)}
-\partial_\beta\partial_\mu F_{\mu\alpha}^{(II)}
- {A+1\over 2} \partial_\mu \partial_\mu F_{\alpha\beta}^{(II)}
+{B\over 2} m^2 F_{\alpha\beta}^{(II)} = 0\,,\\
&&\partial_\alpha\partial_\mu \tilde F_{\mu\beta}^{(III)}
-\partial_\beta\partial_\mu \tilde F_{\mu\alpha}^{(III)}
- {A+1\over 2} \partial_\mu \partial_\mu \tilde F_{\alpha\beta}^{(III)}
+{B\over 2} m^2 \tilde F_{\alpha\beta}^{(III)} = 0\,,\\
&&\partial_\alpha\partial_\mu \tilde F_{\mu\beta}^{(IV)}
-\partial_\beta\partial_\mu \tilde F_{\mu\alpha}^{(IV)}
+ {A-1\over 2} \partial_\mu \partial_\mu \tilde F_{\alpha\beta}^{(IV)}
-{B\over 2} m^2 \tilde F_{\alpha\beta}^{(IV)} = 0\,.
\end{eqnarray}\end{mathletters}
In the Tucker-Hammer case ($A=1, B=2$) we can recover the Proca theory
from (\ref{wth1}):
\begin{equation}
\partial_\alpha \partial_\mu F_{\mu\beta}
-\partial_\beta \partial_\mu F_{\mu\alpha} = m^2 F_{\alpha\beta}
\label{proca1}\,,
\end{equation}
($A_\nu ={1\over m^2}  \partial_\alpha F_{\alpha\nu}$ should be substituted in
$F_{\mu\nu} = \partial_\mu A_\nu -\partial_\nu A_\mu$, and multiplied by
$m^2$).

We also noted~\cite{dvo-wth}
that the massless limit of this theory does {\it not} coincide in full
with  the Maxwell theory, while it contains the latter as a particular
case.  In~\cite{dvo-lg,dv-phtn} we showed that it is possible to define
various massless limits for the Proca-Duffin-Kemmer theory. Another is
the Ogievetski\u{\i}-Polubarinov {\it notoph} (which in the US
literature is called the Kalb-Ramond field), ref.~\cite{Og}. Transverse
components of the AST field can be removed from the corresponding
Lagrangian by means of the ``new gauge transformation" $A_{\mu\nu}
\rightarrow A_{\mu\nu} +\partial_\mu \Lambda_\nu -\partial_\nu
\Lambda_\mu$, with the vector gauge function $\Lambda_\mu$.

The second case is
\begin{equation}
\partial_\alpha\partial_\mu F_{\mu\beta} -\partial_\beta \partial_\mu F_{\mu\alpha}
= [\partial_\mu \partial_\mu - m^2] F_{\alpha\beta}\,.
\end{equation}
So, on the mass shell $[\partial_\mu \partial_\mu - m^2] F_{\alpha\beta} =0$,
and, hence,
\begin{equation}
\partial_\alpha\partial_\mu F_{\mu\beta} -\partial_\beta \partial_\mu
F_{\mu\alpha} = 0\,,\label{str}
\end{equation}
which rather corresponds to the Maxwell-like case. However, if
we calculate dispersion relations for the second case (\ref{str}) it
appears that the equation has solutions even if $m\neq 0$.

The interesting case is $B=8$ and $A=B-1=7$. In this case we can
describe various mass states which are connected
by the relation $m^{\prime^{\,2}} = {4\over 3} m^2$. One can get
\begin{mathletters}
\begin{eqnarray}
&&\partial_\alpha \partial_\mu
F_{\mu\beta} - \partial_\beta \partial_\mu F_{\mu\alpha} =- 3
(\partial_\mu \partial_\mu) F_{\alpha\beta} +4m^2 F_{\alpha\beta}\quad
(\mbox{originated from}\,\,\, P=-1)\,,\\
&&\partial_\alpha \partial_\mu
F_{\mu\beta} - \partial_\beta \partial_\mu F_{\mu\alpha} =+ 4
(\partial_\mu \partial_\mu) F_{\alpha\beta} -4m^2 F_{\alpha\beta} \quad
(\mbox{originated from}\,\,\, P=+1)\,.
\end{eqnarray}
\end{mathletters}
If we consider $\partial_\mu^2 =m^2$, the first
equation will give us only causal solutions with the mass
$m$ which are compatible with the Proca theory;
the second case reduces to (\ref{str}).

Let us consider also $\partial_\mu^2 = \ae m^2$. If one  does not  want to
have neither tachyonic solutions nor old Proca-like solutions
with $m$, one can find another possibility:
set from the first equation $(4-3\ae)m^2 =0$ and, hence, $\ae
={4\over 3}$.  On using this value of $\ae$ in the second equation we
observe that the right-hand side comes to be equal
\begin{equation}
(4\ae - 4) m^2 = {4\over 3} m^2\,, \end{equation} i.e., this is the
compatible solution with the mass $m^{\prime^{\,2}} ={4\over 3} m^2$:
\begin{equation} \partial_\alpha \partial_\mu F_{\mu\beta} -
\partial_\beta \partial_\mu F_{\mu\alpha} = {4\over 3} m^2\,.
\end{equation}

Now we are interested in {\it parity-violating}
equations for antisymmetric tensor fields. We also investigate
the most general mapping of the Weinberg-Tucker-Hammer formulation
to the antisymetric tensor field formulation.
Instead of $\Psi^{(I-IV)}$ we shall try to use now
\begin{equation}
\Psi^{(A)} = \pmatrix{{\bf E} +i{\bf B}\cr
{\bf B} +i{\bf E}\cr} = {1+\gamma^5\over 2} \Psi^{(I)}+
{1-\gamma^5 \over 2} \Psi^{(II)}\,.
\end{equation}
As a result, the equation for the AST fields is
\begin{equation}
\partial_\alpha \partial_\mu F_{\mu\beta}
-\partial_\beta \partial_\mu F_{\mu\alpha}
={1\over 2} (\partial_\mu \partial_\mu) F_{\alpha\beta} +
[-{A\over 2} (\partial_\mu \partial_\mu) + {B\over 2} m^2] \tilde
F_{\alpha\beta}\,.\label{pv1}
\end{equation}
Of course, $\Psi^{(A)^\prime}
=\pmatrix{{\bf B} - i{\bf E}\cr {\bf E} -i{\bf B}\cr} = -i \Psi^{(A)}$,
and the equation is unchanged.
The different choice is
\begin{equation} \Psi^{(B)} = \pmatrix{{\bf E} +i{\bf
B}\cr -{\bf B} -i{\bf E}\cr} = {1+\gamma^5\over 2} \Psi^{(I)}- {1-\gamma^5
\over 2} \Psi^{(II)}\,.
\end{equation}
Thus, one has
\begin{equation}
\partial_\alpha \partial_\mu F_{\mu\beta}
-\partial_\beta \partial_\mu F_{\mu\alpha}
={1\over 2} (\partial_\mu \partial_\mu) F_{\alpha\beta} +
[{A\over 2} (\partial_\mu \partial_\mu)- {B\over 2} m^2] \tilde
F_{\alpha\beta}\,.\label{pv2}
\end{equation}
Of course, one can also
use the dual tensor (${\bf E}^i = -{1\over 2} \epsilon_{ijk} \tilde
F_{jk}$ and ${\bf B}^i =-i\tilde F_{4i}$) and obtain analogous equations:
\begin{mathletters}
\begin{eqnarray}
&&\partial_\alpha \partial_\mu \tilde F_{\mu\beta}
-\partial_\beta \partial_\mu \tilde F_{\mu\alpha}
={1\over 2} (\partial_\mu \partial_\mu) \tilde F_{\alpha\beta} +
[-{A\over 2} (\partial_\mu \partial_\mu) + {B\over 2} m^2]
F_{\alpha\beta}\,,\\
&&\partial_\alpha \partial_\mu \tilde F_{\mu\beta}
-\partial_\beta \partial_\mu \tilde F_{\mu\alpha}
={1\over 2} (\partial_\mu \partial_\mu) \tilde F_{\alpha\beta} +
[{A\over 2} (\partial_\mu \partial_\mu) - {B\over 2} m^2]
F_{\alpha\beta}\,.  \end{eqnarray}
\end{mathletters}
They are connected
with (\ref{pv1},\ref{pv2}) by the dual transformations.

The states corresponding to the new functions $\Psi^{(A)}$,
$\Psi^{(B)}$ etc are {\it not} the parity eigenstates. So, it is not
surprising that we have $F_{\alpha\beta}$ and
its dual $\tilde F_{\alpha\beta}$ in
the same equations. In total we have already eight equations.

One can also consider the most general case
\begin{equation}
\Psi^{(W)} =\pmatrix{aF_{4i} +b \tilde F_{4i} +c \epsilon_{ijk} F_{jk}
+d \epsilon_{ijk} \tilde F_{jk}\cr
eF_{4i} +f \tilde F_{4i} +g \epsilon_{ijk} F_{jk}
+h \epsilon_{ijk} \tilde F_{jk}\cr}\,.
\end{equation}
So, we shall have dynamical equations for $F_{\alpha\beta}$
and $\tilde F_{\alpha\beta}$ with additional parameters $a,b,c,d,\ldots
\in\, {\bf C}$. We have a lot of antisymmetric tensor fields here.
However,
\begin{itemize}

\item
the covariant form preserves if there are some restrictions on the
parameters. Alternatively, we have some additional terms of $\partial_4^2$
or ${\bf \nabla}^2$;

\item
both $F_{\mu\nu}$ and $\tilde F_{\mu\nu}$ are present in the equations;

\item
under the definite choice of $a,b,c,d\ldots$ the
equations can be reduced
to the above equations
for the tensor ${\cal H}_{\mu\nu}$ and its dual:
\begin{equation}
{\cal H}_{\mu\nu} = c_1 F_{\mu\nu} +c_2 \tilde F_{\mu\nu} +
{ c_3\over 2} \epsilon_{\mu\nu\alpha\beta} F_{\alpha\beta}
+{c_4 \over 2} \epsilon_{\mu\nu\alpha\beta} \tilde F_{\alpha\beta}\,;
\end{equation}

\item
the parity properties of $\Psi_W$ are very complicated.
\end{itemize}

\subsection{The Bargmann-Wigner Formalism}

The way for constructing equations of high-spin particles has been given
in~\cite{bw-hs,Lurie}.
However, they claimed
explicitly that they constructed $(2J+1)$ states (the
Weinberg-Tucker-Hammer theory has  essentially $2(2J+1)$
components).  In this subsection we present the
standard Bargmann-Wigner formalism for $J=1$:
\begin{mathletters}
\begin{eqnarray} \left [ i\gamma_\mu \partial_\mu +m \right
]_{\alpha\beta} \Psi_{\beta\gamma} &=& 0\,,\label{bw1}\\ \left [
i\gamma_\mu \partial_\mu +m \right ]_{\gamma\beta} \Psi_{\alpha\beta} &=&
0\,, \label{bw2} \end{eqnarray} \end{mathletters}
If one has
\begin{equation} \Psi_{\left \{ \alpha\beta
\right \} } = (\gamma_\mu R)_{\alpha\beta} A_\mu +
(\sigma_{\mu\nu} R)_{\alpha\beta} F_{\mu\nu}\,,
\end{equation} with
\begin{equation} R = e^{\i\varphi}
\pmatrix{\Theta&0\cr 0&-\Theta\cr}\,\quad
\Theta=\pmatrix{0&-1\cr
1&0\cr}
\end{equation} in the spinorial
representation of $\gamma$-matrices we obtain
the Duffin-Proca-Kemmer equations:
\begin{mathletters}
\begin{eqnarray}
&&\partial_\alpha F_{\alpha\mu} = {m\over 2} A_\mu\,,\\
&& 2m F_{\mu\nu} = \partial_\mu A_\nu - \partial_\nu A_\mu\,.
\end{eqnarray}
\end{mathletters}
(In order to obtain this set one should add the equations
(\ref{bw1},\ref{bw2}) and compare functional coefficients before the
corresponding commutators, see~\cite{Lurie}).  After the corresponding
re-normalization $A_\mu \rightarrow 2m A_\mu$, we obtain the standard
textbook set:
\begin{mathletters} \begin{eqnarray} &&\partial_\alpha
F_{\alpha\mu} = m^2 A_\mu\,,\\ && F_{\mu\nu} = \partial_\mu A_\nu -
\partial_\nu A_\mu\,.  \end{eqnarray} \end{mathletters}
It gives the equation (\ref{proca1})
for the antisymmetric tensor field. How can one obtain other equations
following the Weinberg-Tucker-Hammer approach?

The recipe for the third equation is simple: use, instead of
$(\sigma_{\mu\nu} R) F_{\mu\nu}$, another symmetric matrix $(\gamma^5
\sigma_{\mu\nu} R)  F_{\mu\nu}$, see~\cite{dv-ps} and Appendix C.
And what about the second and the fourth equations? One can modify the
Dirac equation and form the direct product $\Psi_{\alpha\beta} =
\Psi_\alpha \otimes \Psi^c_\beta$, see~\cite{dvo-wig}.  So, I suggest:

\begin{itemize}

\item
by means of specific similarity transformation, see above and~\cite{g1}:
\begin{equation}
[i\gamma_\mu \partial_\mu +m ] \Psi =0 \Rightarrow
[i\gamma_\mu \partial_\mu +m_1 +m_2\gamma_5 ] \Psi =0\,;
\end{equation}

\item
to use the Barut extension too:
\begin{equation}
[i\gamma_\mu \partial_\mu +m ] \Psi =0 \Rightarrow
[i\gamma_\mu \partial_\mu +a{\partial_\mu \partial_\mu\over m} +\ae ] \Psi
=0\,.
\end{equation}
In such a way we can enlarge the set of possible states.

\end{itemize}

\section{Modified Bargmann-Wigner Formalism}

We begin with
\begin{mathletters}
\begin{eqnarray}
\left [ i\gamma_\mu \partial_\mu + a -b \Box + \gamma_5 (c- d\Box )
\right ]_{\alpha\beta} \Psi_{\beta\gamma} &=&0\,,\\
\left [ i\gamma_\mu
\partial_\mu + a -b \Box - \gamma_5 (c- d\Box ) \right ]_{\alpha\beta}
\Psi_{\gamma\beta} &=&0\,,
\end{eqnarray} \end{mathletters}
$\Box$ is the d'Alembertian.
Thus, we obtain the Proca-like equations:
\begin{mathletters}
\begin{eqnarray} &&\partial_\nu A_\lambda - \partial_\lambda A_\nu - 2(a
+b \partial_\mu \partial_\mu ) F_{\nu \lambda} =0\,,\\ &&\partial_\mu
F_{\mu \lambda} = {1\over 2} (a +b \partial_\mu \partial_\mu) A_\lambda +
{1\over 2} (c+ d \partial_\mu \partial_\mu) \tilde A_\lambda\,,
\end{eqnarray} \end{mathletters}
$\tilde A_\lambda$ is the axial-vector potential (analogous to that
used in the Duffin-Kemmer set for $J=0$). Add`itional constraints are:
\begin{mathletters}
\begin{eqnarray}
&&i\partial_\lambda A_\lambda + ( c+d\partial_\mu \partial_\mu) \tilde \phi
=0\,,\\
&&\epsilon_{\mu\lambda\ae\tau} \partial_\mu F_{\lambda\ae} =0\,,\\
&&( c+ d \partial_\mu \partial_\mu ) \phi =0\,.
\end{eqnarray} \end{mathletters}

The spin-0 Duffin-Kemmer equations are:
\begin{mathletters}
\begin{eqnarray}
&&(a+b \partial_\mu \partial_\mu) \phi = 0\,,\\
&&i\partial_\mu \tilde A_\mu  - (a+b\partial_\mu \partial_\mu) \tilde
\phi =0\,,\\
&&(a+b\, \partial_\mu \partial_\mu) \tilde A_\nu + (c+d\,\partial_\mu
\partial_\mu) A_\nu + i (\partial_\nu \tilde \phi) =0\,.
\end{eqnarray}
\end{mathletters}
The additional constraints are:
\begin{mathletters}
\begin{eqnarray}
&&\partial_\mu \phi =0\,\\
&&\partial_\nu \tilde A_\lambda - \partial_\lambda \tilde
A_\nu +2 (c+d\partial_\mu \partial_\mu ) F_{\nu \lambda} = 0\,.
\end{eqnarray}
\end{mathletters}
In such a way the spin states are {\it mixed} through the 4-vector potentials.
For higher-spin equations similar calculations and conclusions
have been reached by M. Moshinsky et al.~\cite{Mosh} and S. Kruglov~\cite{krug}.
After elimination of the 4-vector potentials we obtain
the equation for the AST field of the second rank:
\begin{equation}
\left [ \partial_\mu \partial_\nu F_{\nu\lambda} - \partial_\lambda
\partial_\nu F_{\nu\mu}\right ]   + \left [ (c^2 - a^2) - 2(ab-cd)
\partial_\mu\partial_\mu  + (d^2 -b^2)
(\partial_\mu\partial_\mu)^2 \right ] F_{\mu\lambda} = 0\,,
\end{equation}
which should be compared with our
previous equations which follow from the Weinberg-like formulation.
Just put:
\begin{mathletters}
\begin{eqnarray}
c^2 - a^2 \Rightarrow {-Bm^2 \over 2}\,,&\qquad& c^2 - a^2 \Rightarrow
+{Bm^2 \over 2}\,,\\
-2(ab-cd) \Rightarrow {A-1\over 2}\,,&\qquad&
+2(ab-cd) \Rightarrow {A+1\over 2}\,,\\
b=\pm d\,.&\qquad&
\end{eqnarray}
\end{mathletters}
(The latter condition serves in order to exclude terms $\sim \Box^2$).
Of course, these sets of algebraic equations have solutions in terms $A$
and $B$. We found them and restored the equations:
\begin{equation}
\left [\partial_\mu \partial_\nu F_{\nu\lambda}
-\partial_\lambda \partial_\nu F_{\nu\mu} \right ]
-{A+1\over 2} \partial_\nu\partial_\nu F_{\mu\lambda} + {B\over 2} m^2
F_{\mu\lambda} =0 \end{equation} and
\begin{equation} \left [\partial_\mu
\partial_\nu F_{\nu\lambda} -\partial_\lambda \partial_\nu
F_{\nu\mu}\right ] +{A-1\over 2} \partial_\nu\partial_\nu F_{\mu\lambda} -
{B\over 2} m^2 F_{\mu\lambda} =0 \end{equation}
Thus, the procedure which
we carried out is the following:  {\it The Modified Bargmann-Wigner
formalism} $\rightarrow$ {\it The AST field equations} $\rightarrow$ {\it
The Weinberg-Tucker-Hammer approach}.

The parity violation and the spin mixing are {\it intrinsic} possibilities
of the Proca-like theories. One can go in a different way:
instead of modifying the equations, consider the spin basis rotations.
In the helicity basis we have (see also~\cite{Ber,Grei}, where it was
claimed explicitly that helicity states cannot be parity eigenstates):
\begin{mathletters}
\begin{eqnarray}
&&\epsilon _{\mu }({\bf p},\lambda =+1)={\frac{1}{\sqrt{2}}}{\frac{e^{i\phi }}{
p}}\pmatrix{ 0, {p_x p_z -ip_y p\over \sqrt{p_x^2 +p_y^2}}, {p_y p_z +ip_x
p\over \sqrt{p_x^2 +p_y^2}}, -\sqrt{p_x^2 +p_y^2}}\,, \\
&&\epsilon _{\mu }({\bf p},\lambda =-1)={\frac{1}{\sqrt{2}}}{\frac{e^{-i\phi }
}{p}}\pmatrix{ 0, {-p_x p_z -ip_y p\over \sqrt{p_x^2 +p_y^2}}, {-p_y p_z
+ip_x p\over \sqrt{p_x^2 +p_y^2}}, +\sqrt{p_x^2 +p_y^2}}\,, \\
&&\epsilon _{\mu }({\bf p},\lambda =0)={\frac{1}{m}}\pmatrix{ p, -{E \over p}
p_x, -{E \over p} p_y, -{E \over p} p_z }\,, \\
&&\epsilon _{\mu }({\bf p},\lambda =0_{t})={\frac{1}{m}}\pmatrix{E , -p_x,
-p_y, -p_z }\,.
\end{eqnarray}
\end{mathletters}
and
\begin{mathletters}
\begin{eqnarray}
&&{\bf E}({\bf p},\lambda =+1)=-{\frac{iEp_{z}}{\sqrt{2}pp_{l}}}{\bf p}-{\frac{
E}{\sqrt{2}p_{l}}}\tilde{{\bf p}},\quad {\bf B}({\bf p},\lambda =+1)=-{\frac{
p_{z}}{\sqrt{2}p_{l}}}{\bf p}+{\frac{ip}{\sqrt{2}p_{l}}}\tilde{{\bf p}}, \\
&&{\bf E}({\bf p},\lambda =-1)=+{\frac{iEp_{z}}{\sqrt{2}pp_{r}}}{\bf p}-{\frac{
E}{\sqrt{2}p_{r}}}\tilde{{\bf p}}^{\ast },\quad {\bf B}({\bf p},\lambda
=-1)=-{\frac{p_{z}}{\sqrt{2}p_{r}}}{\bf p}-{\frac{ip}{\sqrt{2}p_{r}}}\tilde{
{\bf p}}^{\ast }, \\
&&{\bf E}({\bf p},\lambda =0)={\frac{im}{p}}{\bf p},\quad {\bf B}({\bf p}
,\lambda =0)=0,
\end{eqnarray}
\end{mathletters}
with $\tilde {\bf p}=\pmatrix{p_y\cr -p_x\cr -ip\cr}$.
See Appendix D for the 4-potentials and fields in the more common
(parity) basis.

In fact, there  are several modifications of the BW formalism. Thanks
to Professor Z. Oziewicz I came to the following set:
\begin{mathletters}
\begin{eqnarray}
\left [ i\gamma_\mu \partial_\mu + \epsilon_1 m_1 +\epsilon_2 m_2 \gamma_5
\right ]_{\alpha\beta} \Psi_{\beta\gamma} &=&0\,,\\
\left [ i\gamma_\mu
\partial_\mu + \epsilon_3 m_1 +\epsilon_4 m_2 \gamma_5 \right ]_{\alpha\beta}
\Psi_{\gamma\beta} &=&0\,,
\end{eqnarray}
\end{mathletters}
where $\epsilon_i$ are the sign operators. So, at first sight, we have 16
possible combinations for the AST fields. First, we come to
\begin{mathletters}\begin{eqnarray}
&&\left [ i\gamma_\mu \partial_\mu + m_1 A_1 + m_2 A_2\gamma_5
\right ]_{\alpha\beta} \left \{ (\gamma_\lambda R)_{\beta\gamma} A_\lambda
+ (\sigma_{\lambda\ae} R)_{\beta\gamma} F_{\lambda\ae }\right
\}+\nonumber\\ &+&\left [ m_1 B_1 +m_2 B_2 \gamma_5 \right ] \left \{
R_{\beta\gamma}\varphi + (\gamma_5 R)_{\beta\gamma} \tilde \phi +(\gamma_5
\gamma_\lambda R)_{\beta\gamma} \tilde A_\lambda\right \}=0\,,\\
&&\left [
i\gamma_\mu \partial_\mu + m_1 A_1 + m_2 A_2\gamma_5 \right
]_{\gamma\beta} \left \{ (\gamma_\lambda R)_{\alpha\beta} A_\lambda +
(\sigma_{\lambda\ae} R)_{\alpha\beta} F_{\lambda\ae}\right \}-\nonumber\\
&-&\left [ m_1 B_1 +m_2 B_2 \gamma_5 \right ] \left \{
R_{\alpha\beta}\varphi +(\gamma_5 R)_{\alpha\beta} \tilde \phi +(\gamma_5
\gamma_\lambda R)_{\alpha\beta} \tilde A_\lambda\right \}=0\,,
\end{eqnarray}\end{mathletters}
where $A_1 = {\epsilon_1 +\epsilon_3 \over 2}$,
$A_2 = {\epsilon_2 +\epsilon_4 \over 2}$,
$B_1 = {\epsilon_1 -\epsilon_3 \over 2}$,
and
$B_2 = {\epsilon_2 -\epsilon_4 \over 2}$.
Thus for spin 1 we have
\begin{mathletters}
\begin{eqnarray} &&\partial_\mu A_\lambda - \partial_\lambda A_\mu + 2m_1 A_1 F_{\mu \lambda}
+im_2 A_2 \epsilon_{\alpha\beta\mu\lambda} F_{\alpha\beta} =0\,,\\
&&\partial_\lambda
F_{\ae \lambda} - {m_1\over 2} A_1 A_\ae -{m_2\over 2} B_2 \tilde
A_\ae=0\,,
\end{eqnarray} \end{mathletters} with constraints
\begin{mathletters}
\begin{eqnarray}
&&-i\partial_\mu A_\mu + 2m_1 B_1 \phi +2m_2 B_2 \tilde \phi=0\,,\\
&&i\epsilon_{\mu\nu\ae\lambda} \partial_\mu F_{\nu\ae}
-m_2 A_2 A_\lambda -m_1 B_1 \tilde A_\lambda =0\,,\\
&&m_1 B_1 \tilde \phi +m_2 B_2 \phi =0\,.
\end{eqnarray} \end{mathletters}
If we remove $A_\lambda$ and $\tilde A_\lambda$ from this set,
we come to the final results for the AST field.

Actually, we have twelve equations (another four equations coincide with
some of those indicated below):
\begin{mathletters} \begin{eqnarray} (I):\quad
&&\partial_\mu A_\lambda -\partial_\lambda A_\mu +2im_2 \tilde
F_{\mu\lambda} =0\,,\\
&&\partial_\lambda F_{\lambda\ae}=0\,;
\end{eqnarray}
\end{mathletters}
\begin{mathletters}
\begin{eqnarray}
(II):\quad
&&\partial_\mu A_\lambda -\partial_\lambda A_\mu +2m_1  F_{\mu\lambda}
=0\,,\\
&&\partial_\lambda F_{\lambda\ae}+{m_1\over 2} A_\ae -
{m_2\over 2} \tilde A_\ae
=0\,; \end{eqnarray} \end{mathletters}
\begin{mathletters}
\begin{eqnarray}
(III):\quad
&&\partial_\mu A_\lambda -\partial_\lambda A_\mu -2m_1  F_{\mu\lambda}
-2im_2 \tilde F_{\mu\lambda}
 =0\,,\\
&&\partial_\lambda F_{\lambda\ae}-{m_1\over 2} A_\ae  =0\,;
\end{eqnarray}
\end{mathletters}
\begin{mathletters}
\begin{eqnarray}
(IV):\quad
&&\partial_\mu A_\lambda -\partial_\lambda A_\mu -
2im_2 \tilde F_{\mu\lambda}
 =0\,,\\
&&\partial_\lambda F_{\lambda\ae}  =0\,;
\end{eqnarray}
\end{mathletters}
\begin{mathletters}
\begin{eqnarray}
(V):\quad
&&\partial_\mu A_\lambda -\partial_\lambda A_\mu -2m_1 F_{\mu\lambda}
 =0\,,\\
&&\partial_\lambda F_{\lambda\ae}  -{m_1 \over 2} A_\ae
+{m_2\over 2} \tilde
A_\ae=0\,; \end{eqnarray} \end{mathletters}
\begin{mathletters}
\begin{eqnarray}
(VI):\quad
&&\partial_\mu A_\lambda -\partial_\lambda A_\mu -2m_1 F_{\mu\lambda}
 =0\,,\\
&&\partial_\lambda F_{\lambda\ae}  -{m_1 \over 2} A_\ae -
{m_2\over 2} \tilde
A_\ae=0\,; \end{eqnarray} \end{mathletters}
\begin{mathletters}
\begin{eqnarray}
(VII):\quad
&&\partial_\mu A_\lambda -\partial_\lambda A_\mu +2m_1 F_{\mu\lambda}+
2im_2 \tilde F_{\mu\lambda} =0\,,\\
&&\partial_\lambda F_{\lambda\ae}+{m_1 \over 2} A_\ae =0\,;
\end{eqnarray}
\end{mathletters}
\begin{mathletters}
\begin{eqnarray}
(VIII):\quad
&&\partial_\mu A_\lambda -\partial_\lambda A_\mu +2m_1  F_{\mu\lambda}
=0\,,\\
&&\partial_\lambda F_{\lambda\ae}+{m_1\over 2} A_\ae +
{m_2\over 2} \tilde A_\ae
=0\,; \end{eqnarray} \end{mathletters}
\begin{mathletters}
\begin{eqnarray}
(IX):\quad
&&\partial_\mu A_\lambda -\partial_\lambda A_\mu  =0\,,\\
&&\partial_\lambda F_{\lambda\ae}+{m_2\over 2} \tilde A_\ae  =0\,;
\end{eqnarray}
\end{mathletters}
\begin{mathletters}
\begin{eqnarray}
(X):\quad
&&\partial_\mu A_\lambda -\partial_\lambda A_\mu +2m_1 F_{\mu\lambda}
-2im_2 \tilde F_{\mu\lambda}
 =0\,,\\
&&\partial_\lambda F_{\lambda\ae} +{m_1 \over 2} A_\ae =0\,;
\end{eqnarray}
\end{mathletters}
\begin{mathletters}
\begin{eqnarray}
(XI):\quad
&&\partial_\mu A_\lambda -\partial_\lambda A_\mu -2m_1 F_{\mu\lambda}
+2im_2 \tilde F_{\mu\lambda}
 =0\,,\\
&&\partial_\lambda F_{\lambda\ae}  -{m_1\over 2} A_\ae =0\,;
\end{eqnarray}
\end{mathletters}
\begin{mathletters}
\begin{eqnarray}
(XII):\quad
&&\partial_\mu A_\lambda -\partial_\lambda A_\mu  =0\,,\\
&&\partial_\lambda F_{\lambda\ae}  -{m_2\over 2} \tilde A_\ae =0\,.
\end{eqnarray}
\end{mathletters}

In the general case we obtain the equation where
both $F_{\mu\nu}$ and
its dual $\tilde F_{\mu\nu}$ present, as above we obtain from the WTH
formalism
\begin{eqnarray} \lefteqn{{m_1 B_1 \over m_1^2 A_1 B_1 -m_2^2
A_2 B_2} [\partial_\mu \partial_\ae F_{\ae\lambda} -\partial_\lambda
\partial_\ae F_{\ae\mu}] +}\\ &+&{im_2 B_2 \over m_1^2 A_1 B_1
-m_2^2 A_2 B_2} [\partial_\mu \partial_\ae \tilde F_{\ae\lambda}
-\partial_\lambda \partial_\ae \tilde F_{\ae\mu}] =m_1 A_1
F_{\mu\lambda} +im_2 A_2 \tilde F_{\mu\lambda}
\end{eqnarray}
One can
go even further. One can use the Barut equations for the BW input. So, we
can get $16\times 16$ combinations (depending on the eigenvalues of the
corresponding sign operators),\footnote{Some of them can coincide each
other.} and we have different eigenvalues of masses due to $\partial_\mu^2
= \ae m^2$.

What is their physical sense? Why do I think that the shown arbitrarieness
of equations for the AST fields is related to 1) spin basis rotations; 2)
the choice of normalization?  In the common-used basis the three
4-potentials have parity eigenvalues $-1$ and one time-like (or spin-0
state), $+1$; the fields ${\bf E}$ and ${\bf B}$ have also definite parity
properties in this basis.  If we transfer to other  basis, e.g., to the
helicity basis we can see that the 4-vector potentials and
the corresponding fields are superpositions of the vector and the
axial-vector.  Of course, they can be expanded in the fields in the
``old" basis. We are going to investigate this surprising fact in other
publications.

\section{Conclusions}
\begin{itemize}

\item
The addition of the Klein-Gordon equation to the $(J,0)\oplus (0,J)$
equations may change physical content even on the free level.
\item
In the $(1/2,0)\oplus (0,1/2)$ representation it is possible to introduce
the {\it parity-violating} frameworks.
\item
We found the mappings between the Weinberg-Tucker-Hammer formalism for
$J=1$ and the AST fields of the 2nd rank of at least eight types. Four
of them include both $F_{\mu\nu}$ and $\tilde F_{\mu\nu}$, which tells us
that the parity violation may occur during  the study of the corresponding
dynamics.
\item
If we want to take into account the $J=1$ solutions with
different parity properties, the Bargmann-Wigner (BW), the Proca and the
Duffin-Kemmer-Petiau (DKP)
formalisms are to be generalized.
\item
We considered the most general
case, introducing eight scalar parameters. In order to have covariant
equations for the AST fields, one should impose constraints on the
corresponding parameters.
\item
It is possible to get solutions with mass
splitting.
\item
We found the 4-potentials and fields in the helicity
basis. They have different parity properties comparing with the standard
(``parity") basis (cf.~\cite{Ber,Grei}).
\item
The discussion induced us to generalize the BW,
the Proca and the Duffin-Kemmer-Petiau formalisms.  Higher-spin equations
may actually describe various spin, mass, helicity and parity states.  The
states of different parity, helicity, and mass may be present in the same
equation.
\item
On the basis of generalizations of the BW formalism, finally,  we obtained
{\it twelve} equations for the AST fields.

\item
A hypothesis was
presented that the obtained results are related to the spin basis
rotations and to the choice of normalization.
\end{itemize}

\section*{Appendix A. Derivation of the Generalized Dirac Equation}

I use the following equation
for the two-component spinor wave function~\cite{Gers}:
\begin{equation}
(E^2 -c^2 {\bf p}^{\,2}) I^{(2)}\Psi =
\left [E I^{(2)} - c {\bf p}\cdot {\bbox \sigma} \right ]
\left [E I^{(2)} + c {\bf p}\cdot {\bbox \sigma} \right ]
\Psi = \mu_2^2 c^4 \Psi \,.\label{G1}
\end{equation}
One can define two-component `right' and `left' wave functions
\begin{equation}
\phi_R= {1\over \mu_1 c} (i\hbar {\partial \over \partial x_0}
-i\hbar {\bbox \sigma}\cdot {\bbox \nabla}) \Psi,\quad\phi_L=\Psi\,
\end{equation}
with an additional mass parameter $\mu_1$.
In such a way we come to the set of equations
\begin{mathletters}
\begin{eqnarray}
&&(i\hbar {\partial \over \partial x_0}
+i\hbar {\bbox \sigma}\cdot {\bbox \nabla}) \phi_R
={\mu_2^2 c\over \mu_1}\phi_L\,,\\
&&(i\hbar {\partial \over \partial x_0}
-i\hbar {\bbox \sigma}\cdot {\bbox \nabla}) \phi_L
=\mu_1 c\phi_R\,,
\end{eqnarray}
\end{mathletters}
which can be written
in the 4-component form:
\begin{eqnarray}
\label{gde}
\pmatrix{i\hbar (\partial/\partial x_0) &
i\hbar {\bbox \sigma}\cdot {\bbox \nabla}\cr
-i\hbar {\bbox \sigma}\cdot {\bbox \nabla}&
-i\hbar (\partial/\partial x_0)}
\pmatrix{\psi_A\cr\psi_B} = {c\over 2}
\pmatrix{(\mu_2^2/\mu_1
+\mu_1)&
(-\mu_2^2/\mu_1 +
\mu_1)\cr
(-\mu_2^2/\mu_1 +
\mu_1)& (\mu_2^2/\mu_1
+\mu_1)\cr}\pmatrix{\psi_A\cr\psi_B\cr}
\end{eqnarray}
for the function $\psi =
column (\phi_R+\phi_L\quad \phi_R - \phi_L )$.
The equation (\ref{gde}) can be written in the covariant form
(as one can see, the standard representation of $\gamma^\mu$ matrices
was used here):
\begin{equation}
\left [ i\gamma_\mu \partial_\mu + {\mu_2^2 c\over \mu_1 \hbar}
{(1-\gamma^5)\over 2} +{\mu_1 c \over \hbar} {(1+\gamma^5)\over 2}
\right ]\psi = 0\,.\label{gde1}
\end{equation}
If $\mu_1=\mu_2$ we can recover the standard Dirac equation.
As noted in ref.~\cite{g1} this procedure can be viewed as simply
changing the representation of $\gamma^\mu$ matrices (unless
$\mu_1\neq 0$, $\mu_2 \neq 0$).

It is interesting that we also can repeat this procedure
for the definition (or even more general):
\begin{equation}
\phi_L= {1\over \mu_3 c} (i\hbar {\partial \over \partial x_0}
+i\hbar {\bbox \sigma}\cdot {\bbox \nabla}) \Psi,\quad\phi_R=\Psi\,
\end{equation}
since in the two-component equation the parity properties of
the two-component spinor are undefined.
The resulting equation is
\begin{equation}
\left [ i\gamma_\mu \partial_\mu + {\mu_4^2 c\over \mu_3 \hbar}
{(1+\gamma^5)\over 2} + {\mu_3 c \over \hbar} {(1-\gamma^5)\over 2}
\right ]\tilde\psi = 0\,.
\end{equation}
The above procedure can be generalized to {\it any} Lorentz group
representation, i.~e., to {\it any} spin.

\bigskip

\section*{Appendix B. Weinberg $2(2J+1)$-component Formalism}

I reviewed the Weinberg
$2(2J+1)$-component formalism
in ref.~\cite{dvo-rmf}.  I found the following
postulates~\cite{Wein}:
\begin{itemize}

\item
The fields transform according to the formula:
\begin{equation}
U [\Lambda, a] \Psi_n (x) U^{-1} [\Lambda, a] = \sum_m D_{nm}
[\Lambda^{-1}] \Psi_m (\Lambda x +a)\,,\label{1} \end{equation}
where $D_{nm} [\Lambda]$ is some representation of $\Lambda$; $x^\mu \rightarrow
\Lambda^\mu_{\quad\nu} \,\,x^\nu +a^\mu$, and $U [\Lambda, a]$ is a
unitary operator.

\item
For $(x-y)$ spacelike one has
\begin{equation}
[\Psi_n (x), \Psi_m (y) ]_\pm =0\,\label{2}
\end{equation}
for fermion and boson fields, respectively.

\item
The interaction Hamiltonian density is said by S. Weinberg to be scalar,
and it is constructed out of the creation and annihilation operators for
the free particles described by the free Hamiltonian $H_0$.

\item
The $S$-matrix is constructed as an integral of the $T$-ordering
product of the interaction Hamiltonians by Dyson's formula.

\end{itemize}

Weinberg wrote: ``In order to discuss theories with parity conservation it
is convenient to use $2(2J+1)$-component fields, like the Dirac field.
These do obey field equations, which can be derived as\ldots consequences
of (\ref{1},\ref{2})."\,\footnote{In the $(2J+1)$ formalism, fields
obey only the Klein-Gordon equation, according to the Weinberg wisdom.}
In such a way he proceeds to form the
$2(2J+1)$-component object
$$\Psi =\pmatrix{\Phi_\sigma\cr \Xi_\sigma\cr}$$
transforming according to the Wigner rules. They are the following ones:
\begin{mathletters}
\begin{eqnarray}
\Phi_\sigma ({\bf p}) &=& \exp (+\Theta \,\hat {\bf p} \cdot {\bf J})
\Phi_\sigma ({\bf 0}) \,,\label{wr1}\\
\Xi_\sigma ({\bf p}) &=& \exp (-\Theta \,\hat {\bf p} \cdot {\bf J})
\Xi_\sigma ({\bf 0}) \,\label{wr2}
\end{eqnarray} \end{mathletters}
from the zero-momentum frame. $\Theta$ is the boost parameter,
$\tanh \,\Theta =\vert {\bf p} \vert/ E$, \,$\hat {\bf p} =
{\bf p}/ \vert {\bf p} \vert$, ${\bf p}$ is the 3-momentum of the particle,
${\bf J}$ is the angular momentum operator.
For a given representation the matrices ${\bf J}$ can be constructed. In
the Dirac case (the $(1/2,0)\oplus (0,1/2)$ representation) ${\bf J} =
{\bbox \sigma}/2$; in the $J=1$ case (the $(1,0)\oplus (0,1)$
representation) we can choose $(J_i)_{jk} = -i\epsilon_{ijk}$, etc. Hence,
we can explicitly calculate (\ref{wr1},\ref{wr2}), see
$J=1/2, 1, 3/2, 2$ cases in the Table (cf. ref.~\cite{WHG}):

\bigskip

\begin{tabular}{ll}
Spin 0 & 1\\
Spin 1/2 & $[ E+m \pm {\bbox\sigma}\cdot {\bf p} ]/[2m (E+m)]^{1/2} $\\
Spin 1 & $[ m(E+m) \pm (E+m) ({\bf J}\cdot {\bf p}) + ({\bf J}\cdot {\bf
p})^2 ]/[m(E+m)]$\\
Spin 3/2 &$ \frac{[-(E+m)(E-5m)\mp (2/3)(E-13m) ({\bf J}\cdot {\bf p})+
4({\bf J}\cdot {\bf p})^2 \pm (8/3) (E+m)^{-1} ({\bf J}\cdot {\bf p})^3
]}{[32m^3 (E+m)]^{1/2}}$\\
Spin 2 &$\frac{[m^2 (E+m)^2 \mp (1/3)(E-4m)(E+m)^2
({\bf J}\cdot {\bf p}) -(1/6) (E-7m) (E+m) ({\bf J}\cdot {\bf p})^2 \pm
(1/3) (E+m) ({\bf J}\cdot {\bf p})^3 + (1/6) ({\bf J}\cdot {\bf p})^4
]}{[m^2 (m+E)^2]}$\\ \end{tabular}

\bigskip

The task is now to obtain relativistic equations for higher spins.
Weinberg uses the following procedure (see also~\cite{Bar-Muz,Novozh}).
Firstly, he defined the scalar matrix
\begin{equation}
\Pi_{\sigma^\prime \sigma}^{(j)} (q) = (-)^{2j} t_{\sigma^\prime
\sigma}^{\quad \mu_1 \mu_2 \ldots \mu_{2j}} q_{\mu_1} q_{\mu_2}\ldots
q_{\mu_{2j}}
\end{equation}
for the $(J,0)$ representation of the Lorentz group ($q_\mu q_\mu =
-m^2$), with the tensor $t$ being defined by [12a,Eqs.(A4-A5)].
Hence,
\begin{equation}
D^{(j)} [\Lambda] \Pi^{(j)} (q) D^{(j)\,\dagger} [\Lambda] = \Pi^{(j)}
(\Lambda q)\label{equ1}
\end{equation}
Since at rest we have $[{\bf J}^{(j)}, \Pi^{(j)} (m)] =0$, then according
to the Schur's lemma $\Pi_{\sigma\sigma^\prime}^{\quad (j)} (m) = m^{2j}
\delta_{\sigma \sigma^\prime}$. After the substitution of $D^{(j)}
[\Lambda]$ in Eq.  (\ref{equ1}) one has
\begin{equation}
\Pi^{(j)} (q) = m^{2j} \exp (2\Theta \,\hat {\bf q}
\cdot {\bf J}^{(j)})\,.  \end{equation}
One can construct the analogous
matrix for the $(0,J)$ representation by the same procedure:
\begin{equation} \overline{\Pi}^{(j)} (q) = m^{2j} \exp (- 2\Theta
\hat{\bf q}\cdot {\bf J}^{(j)}) \,.  \end{equation}
Finally, by  direct
verification one has in the coordinate representation
\begin{mathletters}
\begin{eqnarray} \overline{\Pi}_{\sigma\sigma^\prime} (-i\partial)
\Phi_{\sigma^\prime} =m^{2j} \Xi_\sigma\,,\\ \Pi_{\sigma\sigma^\prime}
(-i\partial) \Xi_{\sigma^\prime} =m^{2j} \Phi_\sigma\,, \end{eqnarray}
\end{mathletters}
if $\Phi_\sigma ({\bf 0})$ and $\Xi_\sigma ({\bf 0})$ are
indistinguishable.\footnote{Later, this fact has been incorporated in the
Ryder book~\cite{Ryder}. Truely speaking, this is an additional postulate.
It is also possible that the zero-momentum-frame $2(2J+1)$-component objects
(the 4-spinor in the $(1/2,0)\oplus (0,1/2)$ representation, the bivector
in the $(1,0)\oplus (0,1)$ representation, etc.) are connected by an
arbitrary phase factor~\cite{Dv-ff}.}

As a result one has
\begin{equation}
[ \gamma^{\mu_1 \mu_2 \ldots \mu_{2j}} \partial_{\mu_1} \partial_{\mu_2}
\ldots \partial_{\mu_{2j}} +m^{2j} ] \Psi (x) = 0\,,
\end{equation}
with the Barut-Muzinich-Williams covariantly-defined
matrices~\cite{Bar-Muz,Sankar}. For the spin-1 they are:
\begin{mathletters} \begin{eqnarray}
&&\gamma_{44} =\pmatrix{0&\openone\cr \openone&0\cr}\,,\quad
\gamma_{i4}=\gamma_{4i} = \pmatrix{0&iJ_i\cr
-iJ_i & 0\cr}\,,\\
&&\gamma^{ij} = \pmatrix{0&\delta_{ij} -J_i J_j - J_j J_i\cr
\delta_{ij} -J_i J_j - J_j J_i & 0\cr}\, .
\end{eqnarray}
\end{mathletters}
Later Sankaranarayanan and Good considered another version of this
theory~\cite{Sankar}. For the $J=1$ case they introduced the
Weaver-Hammer-Good sign operator, ref.~\cite{WHG}, $m^{2} \rightarrow
m^{2}\, (i\partial/\partial t)/E$, which led to the different parity
properties of an antiparticle with respect to a {\it boson} particle.
Next,  Hammer {\it et al}~\cite{TH} introduced another higher-spin
equation. In the spin-1 case it is:
\begin{equation} [\gamma_{\mu\nu}
\partial_\mu \partial_\nu + \partial_\mu \partial_\mu -2m^2 ] \Psi^{(j=1)}
= 0\,.  \end{equation}
In fact, they added the Klein-Gordon equation to the Weinberg equation.

Weinberg considered massless cases too. He claimed that there is no
problem~[12b] to put $m\rightarrow 0$ in propagators and field functions
of the $(J,0)$ and $(0,J)$ fields. {\it But}, there are indeed problems for
the fields of the $(J/2,J/2)$-types, e.g., for the 4-vector potential.
The Weinberg theorem says~[12b,p.B885]: ``A massless particle operator
$a({\bf p},\lambda)$ of helicity $\lambda$ can {\it only} be used to
construct fields which transform according to representations $(A,B)$ such
that $B-A=\lambda$. For instance, a left-circularly polarized photon with
$\lambda =-1$ can be associated with $(1,0)$, $(3/2,1/2)$, $(2,1)$ fields
but {\it not} with the 4-vector potential $(1/2,1/2)$, at least
until we broaden our notion of what we mean by Lorentz transformations".
He indicated that this is a consequence of the non-semi-simple structure
of the little group. In his book~\cite[\S 5.9]{Weinb}, he gave
additional details of what he meant in the above statement. Indeed,
divergent terms of the 4-vector potential ($\lambda = \pm 1$;
under certain choice of the normalization) in the
$m\rightarrow 0$ limit may be removed by {\it gauge transformations},
see Appendix D.

\section*{Appendix C. Additional Fields in the Bargmann-Wigner Formalism}

I noted~\cite{dv-ps} that it is possible to use
another matrices in the expansion of the Bargmann-Wigner symmetric
spinor~\cite{bw-hs,Lurie}. If
\begin{equation}
\Psi_{\{\alpha\beta\}} = (\gamma^\mu R)_{\alpha\beta} (c_a m A_\mu +
c_f  F_\mu) +(\sigma^{\mu\nu} R)_{\alpha\beta} (c_A m\gamma^5
A_{\mu\nu} + c_F F_{\mu\nu})\, ,\label{si}
\end{equation}
we have ``new" Proca equations
\begin{mathletters}
\begin{eqnarray}
&& c_a m (\partial_\mu A_\nu - \partial_\nu A_\mu ) +
c_f (\partial_\mu F_\nu -\partial_\nu F_\mu ) =
ic_A m^2 \epsilon_{\alpha\beta\mu\nu} A_{\alpha\beta} +
2 m c_F F_{\mu\nu} \, \label{pr1} \\
&& c_a m^2 A_\mu + c_f m F_\mu =
-i c_A m \epsilon_{\mu\nu\alpha\beta} \partial_\nu A_{\alpha\beta} -
2 c_F \partial_\nu F_{\mu\nu}\, . \label{pr2}
\end{eqnarray}
\end{mathletters}
Other generalizations of the Bargmann-Wigner formalism have been
presented therein and in ref.~\cite{dvo-wig}.

\section*{Appendix D. Field Functions in the Momentum Space}

The 4-potentials in the momentum space
are~\cite{Novozh,Weinb,dvo-lg,dv-phtn}
(the standard basis is used):
\begin{mathletters} \begin{eqnarray} u^\mu
({\bf p}, +1)= -{N\over \sqrt{2}m}\pmatrix{p_r\cr m+ {p_1 p_r \over
E_p+m}\cr im +{p_2 p_r \over E_p+m}\cr {p_3 p_r \over
E_p+m}\cr}&\quad&,\quad u^\mu ({\bf p}, -1)= {N\over
\sqrt{2}m}\pmatrix{p_l\cr m+ {p_1 p_l \over E_p+m}\cr -im +{p_2 p_l \over
E_p+m}\cr {p_3 p_l \over E_p+m}\cr}\,,\quad\label{vp12}\\ u^\mu ({\bf
p}, 0) &=& {N\over m}\pmatrix{p_3\cr {p_1 p_3 \over E_p+m}\cr {p_2 p_3
\over E_p+m}\cr m + {p_3^2 \over E_p+m}\cr}\,, \quad
u^\mu ({\bf p}, 0_t) = {N \over m} \pmatrix{E_p\cr p_1
\cr p_2\cr p_3\cr}\,,
\label{vp3}
\end{eqnarray}
\end{mathletters}
($p_{r,l} = p_1 \pm ip_2$).
The corresponding fields are:
\begin{mathletters} \begin{eqnarray}
{\bf B}^{(+)} ({\bf p},
+1) &=& -{iN\over 2\sqrt{2}m} \pmatrix{-ip_3 \cr p_3 \cr ip_r\cr} =
+ e^{-i\alpha_{-1}} {\bf B}^{(-)} ({\bf p}, -1 ) \quad,\quad   \label{bp}\\
{\bf B}^{(+)} ({\bf
p}, 0) &=& i{N \over 2m} \pmatrix{p_2 \cr -p_1 \cr 0\cr} =
- e^{-i\alpha_0} {\bf B}^{(-)} ({\bf p}, 0) \quad,\quad \label{bn}\\
{\bf B}^{(+)} ({\bf p}, -1)
&=& {iN \over 2\sqrt{2} m} \pmatrix{ip_3 \cr p_3 \cr -ip_l\cr} =
+ e^{-i\alpha_{+1}} {\bf B}^{(-)} ({\bf p}, +1)
\quad,\quad\label{bm}
\end{eqnarray}
\end{mathletters}
and
\begin{mathletters}
\begin{eqnarray}
{\bf E}^{(+)} ({\bf p}, +1) &=&  -{iN\over 2\sqrt{2}m} \pmatrix{E_p- {p_1
p_r \over E_p +m}\cr iE_p -{p_2 p_r \over E_p+m}\cr -{p_3 p_r \over
E+m}\cr} = + e^{-i\alpha^\prime_{-1}}
{\bf E}^{(-)} ({\bf p}, -1) \quad,\quad\label{ep}\\
{\bf E}^{(+)} ({\bf p}, 0) &=&  {iN\over 2m} \pmatrix{- {p_1 p_3
\over E_p+m}\cr -{p_2 p_3 \over E_p+m}\cr E_p-{p_3^2 \over
E_p+m}\cr} = - e^{-i\alpha^\prime_0} {\bf E}^{(-)} ({\bf p}, 0)
\quad,\quad\label{en}\\
{\bf E}^{(+)} ({\bf p}, -1) &=&  {iN\over
2\sqrt{2}m} \pmatrix{E_p- {p_1 p_l \over E_p+m}\cr -iE_p -{p_2 p_l \over
E_p+m}\cr -{p_3 p_l \over E_p+m}\cr} = + e^{-i\alpha^\prime_{+1}} {\bf
E}^{(-)} ({\bf p}, +1) \quad,\quad\label{em}
\end{eqnarray}
\end{mathletters}
where we denoted a normalization factor appearing in the
definitions of the potentials (and/or in the definitions of the physical
fields through potentials) as $N$
(it is usually to be chosen equal to 1 for 4-potentials).
The signs $(\pm)$ refer to the
positive- and negative- frequency solutions, respectively.

\bigskip
\bigskip

\acknowledgments
I am grateful to Profs. Y. S. Kim, S. I. Kruglov, Z. Oziewicz
and participants of the recent conferences for
useful discussions. I acknowledge the help of A. D'Amore
in English grammatics.

\end{document}